\documentclass[12pt]{article}
\def\be {\begin{equation}}
\def\beq {\begin{equation}}
\def\ee {\end{equation}}
\def\feq {\end{equation}}
\def\ba {\begin{eqnarray}}
\def\ea {\end{eqnarray}}

\def\lb {\label}

\def\vphi {\varphi}

\def\bi {\begin{itemize}}
\def\ei {\end{itemize}}
\begin{document}
\def\bea{\begin{eqnarray}}
\def\eea{\end{eqnarray}}

\title{  General models of Einstein gravity with a non-Newtonian
weak-field limit}
\author{ Mariano Cadoni\footnote{email: mariano.cadoni@ca.infn.it
}
and  Marcello Casula
\\
{\it \small Dipartimento di Fisica, Universit\`a di Cagliari and
INFN, Sezione di Cagliari }\\
{\small Cittadella Universitaria, 09042 Monserrato, Italy}}
\vfill

\maketitle

{\bf Abstract.}
We investigate Einstein theories of gravity, coupled to a scalar field
$\vphi$
and point-like matter, which are characterized by a scalar field-dependent
matter coupling function $e^{H(\vphi)}$. We show that under mild constraints on the
form of the potential for the scalar field, there are a broad class of
Einstein-like gravity models -characterized by the asymptotic behavior of $H$-
which allow for a non-Newtonian weak-field limit with the  gravitational potential
behaving for
large distances  as $\ln r$. The
Newtonian term $GM/r$ appears only as sub-leading. We point out
that this behavior is also shared by gravity models described by $f(R)$
Lagrangians. The relevance of our results  for the building of
infrared modified theories of gravity and for modified Newtonian
dynamics is also discussed.

\section{Introduction}

In the last decade there has been  growing interest for gravitational
models
with large-distance deviations from standard, Einsteinian and Newtonian
gravity \cite{Dvali:2000hr,ArkaniHamed:2003uy,Milgrom:1983ca,Milgrom:1992hr,deDiego:2008hd,
Bruneton:2007si,Famaey:2006iq,
Trencevski:2006hp,Bertolami:2004hh,Capozziello:2006ph}. The main reason behind this interest is the hope that
the accelerated expansion of the universe and rotational curves of
the galaxies could be explained by large distance modifications of
our present  understanding of gravity  and inertia,
without postulating non-baryonic forms of matter such as dark matter
and
dark energy \cite{Trimble:ds,Kunz:2006ca}.  One of the most promising  possibilities is represented,
in the context of the
brane-world scenario, by the Dvali-Garbadze-Porrati (DGP) model \cite{Dvali:2000hr}.
In the DGP model the gravitational interaction is described by
five-dimensional Einstein gravity and the usual four-dimensional (4D)  Newtonian potential is generated by a 4D scalar curvature term in
the world-volume brane action.

The idea of modifying Newton's law
and/or
Einstein gravity at large distances  is not new. Since galactic rotational curves
were found to be inconsistent with the distribution of luminous
matter, such alternative theories of gravity have been proposed and
pushed forward \cite{Milgrom:1983ca,Milgrom:1992hr,Bruneton:2007si}.
On the other hand, from the experimental side,  there is  plenty of
room
for such infrared modifications of gravity; our experimental knowledge
of gravity conforms with general relativity but is limited to
distances between say $~ 10^{-3}cm - 100 MPc$.

A quite generic prediction of infrared-modified theories of gravity
is the presence of large-distance logarithmic corrections to the
Newtonian potential. In the DGP model, a $\ln(r/L)$
term, where $L$ is a cross-over length,
appears  as correction of the Newtonian potential  \cite{Dvali:2000hr,Cadoni:2008fb}.
In models that  modify Newtonian
dynamics to account for galactic rotational curves the  logarithmic
corrections
dominates for small accelerations, of order $a_{0}\sim 10^{-29}cm^{-1}$ (in natural units) \cite{Milgrom:1983ca,Milgrom:1992hr}.

The search for
Einstein-like theories of gravity that allow for a non-Newtonian
weak-field limit with large-distance logarithmic correction  is therefore of great interest. Such models can be considered as an
effective descriptions of more general  theories of gravity, such as
those emerging in the brane-world scenario, which modify the
gravitational interaction at large distance.

It is not easy to build metric, Einstein-like, theories of
gravity that allow
for a non-Newtonian weak-field limit. Typically one has to introduce
additional fields (scalars or vectors) and  add to the
Einstein-Hilbert action additional terms. Alternatively, one may
consider higher-curvature terms in the gravitational action, e.g. in
the form of $f(R)$ theories. In both cases the theory
is highly constrained by several phenomenological requirements:
existence of a  Newtonian potential term  at  short distances,
absence of a fifth force,  correct predictions for light bending and
so on (see e.g. Ref. \cite{Bruneton:2007si}).

A simple theory of Einstein gravity coupled
to a scalar field with an exponential potential that allows for a
$\ln(r/L)$
weak field limit has been proposed in Ref. \cite{Cadoni:2003nz}. The proposed gravity
theory requires a particular form of both the scalar field potential
and  the coupling function between scalar field and matter.

In this paper we show that the existence of a  $\ln(r/L)$
weak-field limit is not restricted to the particular model of Ref. \cite{Cadoni:2003nz}, but is a quite generic feature of a broad class of models of
gravity coupled with a scalar field. We require that the model
satisfies the following conditions: \\
$a)$ Existence of a consistent
non-relativistic, static, weak-field limit of the gravitational field
equations.\\
$b)$ At large distance ($r\gg L$)  the gravitational
potential is dominated by the $\ln(r/L)$ term.\\
$c)$ Presence of a
sub-leading Newtonian term $-GM/r$.\\
$d)$ The short distance deviations from Newtonian behavior are at
most of order $r^{-\gamma}$ with $\gamma\geq 2$.\\
We show that virtually all models of  Einstein gravity  coupled with a
scalar field satisfies the above  conditions if an appropriate
coupling function
 between the scalar field and matter is introduced. As an
illustration of our procedure we discuss in detail the case of a
scalar field with a power-law potential.
 The models can be also
rewritten, in the Jordan  frame, as a scalar-tensor theory of gravity
and also reformulated as a theory of
gravity with an $f(R)$ Lagrangian.

The structure of the paper is as follows. In section 2 we present
our model. In section 3 we discuss the weak-field limit. In section
4 we  derive the general solution. In section 5 we illustrate our
general procedure for the particular case of a scalar field with a
power-law potential.  The relevance of our models
for describing the infrared modifications of the gravitational
interaction is discussed  in section 6. In section 7 we formulate our
model in the Jordan frame and we stress its equivalence with a
$f(R)$ theory of gravity. Finally, in section 8 we state our
conclusions.

\section{The model }
Let us consider a system of two point-particles of mass $M$ (the
source) and $m$ (test particle),
with $M\gg m$ interacting with the gravitational field $g_{\mu\nu}$ and
a scalar field $\varphi$.
The gravitational field is described by the Einstein-Hilbert
action.
The scalar field is self-interacting, with potential $V(\varphi,L)$,
and its
interaction with the point-particles is characterized by a coupling
function
$e^{H(\varphi,L)}$.
Both the potential  and the coupling function
depend not only on $\varphi$ but also on a real parameter $L$
with dimensions of a length (we use natural units).
For a generic, $V(\vphi)\neq \lambda \vphi^{4}$, potential the presence of such
a parameter
is necessary for purely dimensional
reasons,
because the scalar field $\varphi$ has dimension of a
(length)$^{-1}$. We will show later in detail that this parameter has the
physical meaning of the length-scale, above which the non-Newtonian
behavior of our gravitational model become relevant.

The classical dynamics of the system is described by the
Einstein-like action
\beq\lb{action}
S=\int d^{4}x \sqrt{-g}\left[\frac{1}{16\pi G} R -
\partial_{\nu}\varphi\partial^{\nu}\varphi
-V\right]- e^{H}\left(M\int ds_{(M)}+ m\int ds_{(m)}\right),
\feq
where $ds_{(M,m)}$ are respectively the line elements of the source and
test particle  $ds_{(M,m)}= dt [(dx_{(M,m)}^{\mu}/dt)
(dx_{(M,m)}^{\nu}/dt)g_{\mu\nu}]^{1/2}$ and we are using for the
metric  a
signature $(-1,1,1,1)$.

In the limit $M\gg m$  the contribution
of the test particle to the stress-energy tensor and to the equation
for the scalar field
can be neglected. Moreover, we are only interested  in the motion of
the
test particle.
The resulting equation of motion stemming from the action
(\ref{action}) are,
\bea\lb{fe}
&&R_{\mu\nu}-{1\over 2} g_{\mu\nu} R= 8\pi G\left[
T^{(\varphi)}_{\mu\nu}+e^{H}¥T^{(M)}_{\mu\nu}\right],\nonumber\\
&&2\nabla^{2}\varphi- \frac{\partial V}{\partial \varphi}=M H'e^{H}¥
\int d\tau
{\delta^{4}(x^{\nu}-x_{(M)}^{\nu}(\tau))\over\sqrt{-g}},\\
&&\frac{d^{2}x_{(m)}¥^{\mu}}
{d\tau^{2}¥}+\tilde\Gamma^{\mu}_{\rho\sigma}\frac{dx_{(m)}¥^{\rho}}{
d\tau}
\frac{dx_{(m)}¥^{\sigma}}{ d\tau}=0,\nonumber
\eea
where the prime denotes derivation with respect to $\varphi$ and
$T^{(\varphi)}_{\mu\nu}$,  $T^{(M)}_{\mu\nu}$ are the
stress-energy tensors
for the scalar field  and for the source of mass $M$,respectively,  and
$\tilde\Gamma$ is a  $\varphi$-dependent connection:
\bea\lb{Stress}
T^{(\varphi)}_{\mu\nu}&=& 2
\partial_{\mu}\varphi\partial_{\nu}\varphi-g_{\mu\nu}\left[(\partial
\varphi)^{2}+V\right],\nonumber\\
T^{(M)}_{\mu\nu}&=&M\int d\tau
u_{\mu}u_{\nu}¥\frac{\delta^{4}(x¥^{\alpha}-x_{(M)}¥^{\alpha}(\tau))}{\sqrt{-g}},
\\
\tilde\Gamma^{\mu}_{\rho\sigma}&=&\Gamma^{\mu}_{\rho\sigma}+\frac{1}
{2}\left(\partial _{\rho}H\delta^{\mu}_{\sigma}+\partial
_{\sigma}H\delta^{\mu}_{\rho}-2g_{\rho\sigma}g^{\mu\gamma}
\partial_{\gamma}H\right)\nonumber.
\eea
In the previous equations $u_{\mu}$ is the
four-velocity of the source.
\section{The weak-field  limit}
We are interested in  the usual weak-field, non-relativistic, static
limit of the field equations (\ref{fe}). We expand the metric near a
flat background, $g_{\mu\nu}=
\eta_{\mu\nu}+h_{\mu\nu}$ with $h_{\mu\nu}\ll 1$ and we  consider
field configurations
depending only on the spatial coordinates $ x^{i},\,
i=1,2,3$. In  the non-relativistic limit, when the velocity of the
particles are $ \ll 1$ and $|T_{ij}|\ll|T_{00}|$,  the field equations
(\ref{fe}) give,
\bea\lb{nl}
\bar\nabla^{2}\Phi&=&4\pi G\left[(\bar\nabla
\varphi)^{2}+V\right]+ H'' (\nabla \varphi)^{2}+\frac{1}{2}H'V'+
e^{H}\left( 4\pi G +\frac{H'^{2}}{2}\right)\tilde
T^{(M)}_{00}\nonumber\\
\bar\nabla^{2}\varphi&=& \frac{1}{2} V' +\frac{1}{2} e^{H} H'
\tilde T^{(M)}_{00}\\
{d^{2}\bar x_{(m)}¥\over dt^{2}}&=&-\bar\nabla\Phi \nonumber,
\eea
where $\Phi=-h_{00}/2+H$, $\tilde T^{(M)}_{00}= M\delta^{3}(\bar x-
\bar
x_{(M)})$. The bar indicates three-dimensional vectorial quantities,
and   the differential operators are calculated with respect to the
three-dimensional Euclidean metric.

From the third  Equation in  (\ref{nl}) it is evident that in the
weak-field limit the field
$\Phi$ represents the
potential that determines the force acting on the test particle.
The usual weak-field Newtonian limit can be trivially recovered
setting
in Eqs. (\ref{nl}) $H=V=0$ and picking the $\varphi=0$ solution for
the scalar field equation.
In the following we will only consider spherical symmetric
solution  to the Eqs. (\ref{nl}). They  can be found placing the
source particle of mass $M$ at the origin of the coordinate system and
using spherical coordinates $(r,\theta, \omega)$.

\section{The general solution}
Our goal is to find a general class of models described by the action
(\ref{action}) and parametrized by a particular, albeit general, from
of the functions $V(\vphi,L)$, $H(\vphi,L)$,  whose solutions
satisfy the
following conditions:\\
$a)$ Existence of a consistent
non-relativistic, static, weak field limit given by Eq. (\ref{nl}).\\
$b)$ At large distance ($r\gg L$)  the solutions for $\Phi$ are
dominated by
a $\ln(r/L)$ term .\\
$c)$ They  must allow for a sub-leading Newtonian term $-GM/r$.\\
$d)$ The short distance deviations  of $\Phi$ from the Newtonian
behavior are at
most of order $r^{-\gamma}$ with $\gamma\geq 2$.

Obviously, these conditions constraint the form of the
scalar potential $V$ and on the coupling function $H$.
Consistency  of the non-relativistic approximation
requires that  the stress-energy tensor satisfies the following condition:
$|T_{ij}|\ll|T_{00}|$.
In the case of the source  of the gravitational field the previous conditions are
 satisfied if one  considers a point particle moving with
speed much lesser then the speed of light. On the other hand for the
stress-energy tensor  of the scalar field in Eq. (\ref{Stress})  we have:
\beq
T^{(\vphi)}_{00}= (\partial_{r}\vphi)^{2}+V,\quad  
T^{(\vphi)\mu}_{\mu}=-4 V- 2
(\partial_{r}\vphi)^{2}.
\feq
It follows that the  conditions for the validity of the non
relativistic approximation
are consistent with the
weak-field equations (\ref{nl}) only if  $V$ and
$(\partial_{r}\vphi)^{2}$ do not affect the Newtonian term in the
potential $\Phi$.
This together with condition $d)$ above imply    the large $r$,
$r\gg L$
behavior:
\beq\lb{e2}
V(\vphi)\sim (\partial_{r}\vphi)^{2}\,\sim {\cal O}(r^{-\beta}),
\quad \beta\geq
4.
\feq

Condition $c)$ requires the presence of the term $- {MG}/{r}$ in
the solution for $\Phi$. In view of the first equation in (\ref{nl})
this implies $(4\pi G+ H'^2)e^H|_{r=0}=4\pi G.$
The simplest solution to this equation is given by
\beq\lb{e4}
H(r=0)=H'(r=0)=0.
\feq
Our  next and last task is to fulfil condition $b)$. Using (\ref{e4})
and the equations of motion for the scalar field $\varphi$, the first equation in (\ref{nl})
becomes
\beq\lb{e6}
\bar \nabla^2 \Phi=\bar \nabla^2 H+ 4\pi G T^{(M)}_{00}.
\feq
In order to have a solution for  the gravitational potential $\Phi$
satisfying at the same time condition $d)$, we must have for $r\gg L$
\beq\lb {e8}
\bar \nabla^2 H= \frac{C}{r^2}+{\cal O}(r^{-4}),
\feq
where C is an arbitrary constant. Obviously Eq. (\ref{e8}) determines
the dependence of $H$ on $r$. In order to find the dependence of the
coupling function $H$ on $\vphi$, we need to solve first the
equation of motion for $\vphi(r)$.
The $r\gg L$ asymptotic form of the  solution  of Eq. (\ref{e8}) is
\beq\lb {e18}
    H= C\ln\frac{r}{L}+ C_1+{\cal O}(r^{-2}),
\feq
where $C_1$ is an arbitrary constant.
Notice that in principle we can use Eq. (\ref{e6}) to describe an
arbitrary deviation
(not necessarily logarithmic) of the gravitational potential from its
Newtonian behavior. Obviously in this case Eq. (\ref{e8}) as to be
modified accordingly.

We have now characterized completely the general model we are looking
for.
Every theory of gravity coupled with a scalar described by the action
(\ref{action}),
will allow in the non-relativistic, static, weak-field limit for  a
solution
\beq\lb{e10}
\Phi= C\ln\frac{r}{L}-\frac{GM}{r}+{\cal O}(r^{-2}),
\feq
if the coupling function satisfies Eqs. (\ref{e4}),(\ref{e18}).
For instance a simple solution of Eqs. (\ref{e4}),(\ref{e18}) is
 \beq\lb{e11}
H= \frac{C}{2}\ln(1+\frac{r^2}{L^2}).
\feq

The form of the potential $V(\vphi)$ is only constrained by  Eq.
(\ref{e2}), which is necessary for the consistency of the
non-relativistic, weak-field approximation.
An other physical condition  that has to be imposed on the form of the
potential is stability, i.e. $V(\vphi)$ must be bounded  from  below
and eventually it must allow for a local minimum. Excitations near
this minimum must have enough heavy mass to be compatible with
particle physics phenomenology.
\section{ Scalar field with a power-law potential}
As a  simple example and illustration of our general approach we
consider the case of a scalar field with a power-law potential
\beq\lb{e11a}
V= N L^{-4}(\vphi L)^{(2-2/\alpha)},
\feq
where $N$ and $\alpha$ are arbitrary dimensionless parameters.
The weak-field equations  (\ref{nl}) for $\vphi$  are readily solved to
give
\beq\lb{e12}
\vphi= D L^{-1}\left(\frac{r} {L}\right)^{\alpha},
\feq
where $D= \left([N(\alpha -1)]/[\alpha^2(\alpha+1)\right]^{\alpha/2}$.
Using solution (\ref{e12}) one easily find that the consistency
conditions (\ref{e2}) are satisfied if $\alpha<-1$.

 The coupling function $H$ is constrained by Eqs. (\ref{e4}),(\ref{e18}).
 A simple solutions is given by Eq. (\ref{e11}), which in view of
 Eq. (\ref{e12}) becomes:
 \beq\lb{eq9}
 H(\vphi)=\frac{C}{2}\ln\left[1+\left(\frac{L\vphi}{D}\right)^\frac{2}{\alpha}\right].
 \feq
For $\alpha<-1$ the potential $ V(\vphi)$ is bounded from below in
the range $ 0< \vphi <\infty$, corresponding to the physical range
$0< r <\infty $ of the radial coordinate $r$. Thus the system is
stable and in principle one can also deform the potential in such a
way that $V(\vphi)$ has  local minimum at $\vphi=\vphi_0$, with
$V''(\vphi_0)=m^2_\vphi$. The parameters in the potential have to be
chosen in  such a way that the mass $m_\vphi$ of the local excitation is
heavy enough to be compatible with the experimental results about
the absence of long range scalar forces.

\section{ Large-distance modifications of gravity}
The gravity model (\ref{action}) can be used  both as an
effective description of infrared modification of gravity, e.g. in the
brane world scenario, or also as an Einstein-like description of
Newtonian modified dynamics.

In the first case the scalar field $\vphi$ could represent  an
effective 4D parametrization of those effects that modifies gravity
at large distances ( for instance the embedding of our 3-brane in a
higher dimensional world).
Apart from the potential $V(\vphi)$, which does not determine the
infrared modifications of gravity,   the model is parametrized by a
coupling function $H$ which depends on a length scale $L$ and a
coupling constant $C$.
$H$ is zero together with its first derivatives at $r=0$ and stays
almost constant (and vanishing) for $r\ll L$. In this regime the
gravitational potential is given by the Newtonian expression  and the
theory reduces to  General relativity.
For $r\gg L$, $H\sim \ln(r/L)$,  the gravitational potential is
dominated by a $\ln r$ term and the theory deviates from general
relativity, still giving an effective metric description of the
gravitational interaction.

In the second case, one can  use our model to explain the rotation curves
of the galaxies without postulating the presence of dark matter. To
this end one introduces a constant acceleration $a_0\sim 10^{-29}cm$
such that for $a\gg a_0$ one recovers standard Newtonian dynamics. In
fact, in this case $\sqrt{GM}/r\gg a_0$ and the Newtonian term in the
potential (\ref{e10}) dominates over the logarithmic term. On the
other hand for
$a\sim a_0$  the leading logarithmic term dominates the gravitational
potential (\ref{e10}) and the rotation curves of the galaxies can be
explain using the gravitational potential (\ref{e10}) and identifying
the constant $C$ as

\beq\lb{e25}
C=\left(\frac{M}{\sigma}\right)^{1/2}¥,
\feq
where $M$ is the mass of the  galaxy and $\sigma$ is an universal
constant, which can be determined using  the Tully-Fisher law \cite{Tully:1977fu,Cadoni:2003nz}.

\section{Jordan frame and equivalence with $f(R)$ theories}

The gravitational model described by the action (\ref{action}) is
equivalent to a scalar-tensor theory of gravity. It is  a
scalar-tensor theory of gravity written in the Einstein frame. The
transition from the Einstein to the Jordan frame is given by the Weyl
transformation,
\beq\lb{e15}
g_{\mu\nu}=e^{-2H}\hat g_{\mu\nu},
\feq
where $\hat g_{\mu\nu}$ is the metric in the Jordan frame.
In this  frame the action (\ref{action}) becomes
\bea\lb{action1}
S&=&\int d^{4}x \sqrt{-\hat g}e^{-2H}¥\left[\frac{1}{16\pi G} R(\hat
g) +
\left(\frac{3}{8\pi G} H'-1\right)\partial_{\nu}\varphi\partial^{\nu}\varphi
-e^{-2H}¥V\right]+\nonumber\\
&-& \left(M\int ds_{M}+ m\int ds_{m}\right).
\eea

In the Einstein frame theories described by the action (\ref{action})
lead to a breakdown of the equivalence principle at galactic scales
\cite{Cadoni:2003nz}.
Conversely in the  Jordan frame the particles do not couple to the
scalar field.
The geodesic equation depends only on the metric $\hat g_{\mu\nu}$.
On the other hand in the Jordan frame the theory has the usual
weaknesses of scalar-tensors theories of gravity: Newton constant
depends on the coordinates and light bending is not correctly
reproduced (see e.g. Ref. \cite{Bruneton:2007si}).

One can also  consider the weak-field limit of the field
equations stemming from the action (\ref{action1}) along the lines
described in the previous sections.
Because now  the geodesic equation for
the test particle depends only on  the metric $\hat g_{\mu\nu }$, we set $\hat g_{\mu\nu }=\eta_{\mu\nu}+\hat h_{\mu\nu}$
and $\Phi=-\hat h_{00}/2$.
In the non-relativistic, weak-field limit we can approximate $e^{\pm2H
}=1\pm 2H$ and  the   equation  of motion for $\Phi$, $\vphi$ and the
geodesic equation coincide  with (\ref{nl}) upon use of Eqs.
(\ref{e2}),(\ref{e4}).
Hence the solutions (\ref{e18}),(\ref{e10}),(\ref{e11}) found in Sect.
4 are also solution of the theory in the Jordan frame.

It is  well known that gravity models whose Lagrangians are given by
functions of the scalar curvature $f(R)$  are on-shell equivalent to
Einstein gravity coupled to a scalar field (see e.g. Ref. \cite{Bruneton:2007si}).
General models
may also involve a an arbitrary coupling function $e^{H(\vphi)}$
\cite{Bruneton:2007si} and be therefore equivalent to
the model described  by the action (\ref{action}). The potential
$V(\vphi)$ for the scalar field is determined by the the function
$f(R)$. Our characterization of gravity models with non-Newtonian
weak-field limit can be therefore easily extended  to the case of
gravity theories  with $f(R)$ Lagrangians.

The existence of non-Newtonian solutions seems a quite generic
feature of $f(R)$ theories. In fact the potential $V$
is determined entirely  by the  function $f$ and  the requirement for the
existence of non-Newtonian solutions constrain very weakly the form
of $V$. Also in this case the existence of the non-Newtonian
weak-field limit is related to the presence of a non trivial coupling
function $e^{H}$.

\section{ Conclusion}
In this paper we have show that the existence of a non-Newtonian
weak-field limit  in which the gravitational potential has a large
distance $\ln(r/L)$ leading term is a quite generic feature of a broad
class of models of Einstein 
gravity coupled with a scalar field and  of theories  of
gravity with an $f(R)$ lagrangian,  when a scalar field-dependent
matter coupling function is introduced.
This class of models has been selected by enforcing rather general
physical conditions on the form of the non-relativistic gravitational
potential.

The  gravity  models proposed and investigated  in this paper
may be very useful to give an effective
description for large-distance deviations of the gravitational
interaction from its GR behavior, e.g in brane-world scenarios.
Another interesting field of application is their use to describe
modifications of Newtonian dynamics at small accelerations, e.g. for
explaining rotation curves of galaxies.

The most sensible and critical point of our approach is the physical
origin  of  the coupling function $e^{H(\vphi)}$. Its
presence is quite natural in scalar-tensor theories of gravity, e.g.
Brans-Dicke theory and related to the choice of what may be called
the physical Weyl frame \cite{Faraoni:1999hp,Capozziello:1996xg}.
But its physical interpretation is far from
being clear. One possibility is that it gives a local effective
description of non-local effects. For instance, in brane-world
scenarios it may be considered as a local
parametrization of nonlocal effects generated by the embedding of
the 3D brane in the higher dimensional spacetime.


\begin{thebibliography}{99}
\bibitem{Dvali:2000hr}
  G.~R.~Dvali, G.~Gabadadze and M.~Porrati,
  Phys.\ Lett.\  B {\bf 485} (2000) 208
  [arXiv:hep-th/0005016].

\bibitem{ArkaniHamed:2003uy}
  N.~Arkani-Hamed, H.~C.~Cheng, M.~A.~Luty and S.~Mukohyama,
gravity,''
  JHEP {\bf 0405} (2004) 074
  [arXiv:hep-th/0312099].

\bibitem{Milgrom:1983ca}
M.~Milgrom,
Astrophys.\ J.\  {\bf 270} (1983) 365.

\bibitem{Milgrom:1992hr}
M.~Milgrom,
Annals Phys.\  {\bf 229} (1994) 384
[arXiv:astro-ph/9303012].
%
\bibitem{deDiego:2008hd}
  J.~A.~de Diego,
  arXiv:0807.0617 [physics.space-ph].


\bibitem{Bruneton:2007si}
  J.~P.~Bruneton and G.~Esposito-Farese,
  Phys.\ Rev.\  D {\bf 76} (2007) 124012
  [Erratum-ibid.\  D {\bf 76} (2007) 129902]
  [arXiv:0705.4043 [gr-qc]].

\bibitem{Famaey:2006iq}
  B.~Famaey, G.~Gentile, J.~P.~Bruneton and H.~S.~Zhao,
  Phys.\ Rev.\  D {\bf 75} (2007) 063002
  [arXiv:astro-ph/0611132].

\bibitem{Trencevski:2006hp}
  K.~Trencevski and E.~G.~Celakoska,
  arXiv:gr-qc/0604068.

\bibitem{Bertolami:2004hh}
  O.~Bertolami and J.~Paramos,
  arXiv:gr-qc/0411020.

\bibitem{Capozziello:2006ph}
  S.~Capozziello, V.~F.~Cardone and A.~Troisi,
  Mon.\ Not.\ Roy.\ Astron.\ Soc.\  {\bf 375}, 1423 (2007)
  [arXiv:astro-ph/0603522].
\bibitem{Trimble:ds}
V.~Trimble,
Ann.\ Rev.\ Astron.\ Astrophys.\  {\bf 25} (1987) 425.

\bibitem{Kunz:2006ca}
  M.~Kunz and D.~Sapone,
  Phys.\ Rev.\ Lett.\  {\bf 98} (2007) 121301
  [arXiv:astro-ph/0612452].

\bibitem{Cadoni:2008fb}
  M.~Cadoni and P.~Pani,
  arXiv:0812.3010 [hep-th].


\bibitem{Cadoni:2003nz}
  M.~Cadoni,
  Gen.\ Rel.\ Grav.\  {\bf 36} (2004) 2681
  [arXiv:gr-qc/0312054].
\bibitem{Tully:1977fu}
  R.~B.~Tully and J.~R.~Fisher,
  Astron.\ Astrophys.\  {\bf 54} (1977) 661.

\bibitem{Faraoni:1999hp}
  V.~Faraoni and E.~Gunzig,
  Int.\ J.\ Theor.\ Phys.\  {\bf 38} (1999) 217
  [arXiv:astro-ph/9910176].
\bibitem{Capozziello:1996xg}
  S.~Capozziello, R.~de Ritis and A.~A.~Marino,
  Class.\ Quant.\ Grav.\  {\bf 14} (1997) 3243
  [arXiv:gr-qc/9612053].


\end{thebibliography}
\end{document}